\documentclass[onecolumn,
preprint,
preprintnumbers,
nofootinbib,
amsmath,amssymb,
aps,
prd,
]{revtex4-2}

\usepackage{graphicx}
\usepackage{dcolumn}
\usepackage{bm}
\usepackage{amssymb}
\usepackage{hyperref}
\usepackage{multirow}
\usepackage{slashed}
\DeclareUnicodeCharacter{2212}{\textendash}
\def\be{\begin{equation}}
\def\ee{\end{equation}}
\def\bea{\begin{eqnarray}}
\def\eea{\end{eqnarray}}

\usepackage{makecell}

\usepackage{comment}
\renewcommand{\comment}[1]{}

\usepackage{xcolor}

\makeatletter
\newcommand{\Biggg}{\bBigg@{3.5}}
\makeatother

\begin{document}

\begin{flushright}
TUM-HEP-1594/26\\
NORDITA 2026-027\\
\vspace*{1cm}
\end{flushright}

\title{Dark Acoustic Oscillations and the Hubble Tension}
\author{Mathias Garny$^1$}\email{mathias.garny@tum.de}
\author{Florian Niedermann${}^2$}\email{florian.niedermann@su.se}
\author{Martin S. Sloth${}^3$}\email{sloth@sdu.dk}
\affiliation{%
\small ${}^1$Physik Department T31, School of Natural Sciences, Technische Universit\"at M\"unchen \\
James-Franck-Stra\ss e 1, D-85748 Garching, Germany\\
${}^2$Nordita, KTH Royal Institute of Technology and Stockholm University\\
Hannes Alfv\'ens v\"ag 12, SE-106 91 Stockholm, Sweden
\\
${}^3$ Universe-Origins, University of Southern Denmark, Campusvej 55, 5230 Odense M, Denmark}%

\begin{abstract}
The Hubble tension and the recently reported anomaly in  data from the Dark Energy Spectroscopic Instrument (DESI) are considered to pose serious challenges to the standard $\Lambda$CDM model of cosmology. 
In this work, we show that resolving the Hubble tension with a scenario featuring dark radiation–matter decoupling (DRMD) predicts the presence of dark acoustic oscillations (DAO) similar in scale to baryon acoustic oscillations (BAO). Using an inference independent of large-scale structure data, relying only on Planck measurements of the cosmic microwave background and SH$0$ES-calibrated supernova data, we find evidence for a DAO signal with drag-horizon scale $r_{d,\mathrm{DAO}} \in[54,65]\,\mathrm{Mpc}/h$ ($68\%\,\mathrm{C.I.}$) and amplitude $A_\mathrm{DAO} \in [0.02,0.05]$ ($68\%\,\mathrm{C.I.}$). These predictions provide a concrete target for current and upcoming large-scale structure surveys, including DESI, Euclid, and the Roman Space Telescope.
Remarkably, the predicted DAO properties are consistent with those required to explain the DESI anomaly, offering both an alternative to evolving dark energy and a preliminary validation of the relevance of a dark radiation–matter decoupling scenario for addressing the Hubble tension.
\end{abstract}

\maketitle 

\newpage

\section{Introduction}

Observational challenges to $\Lambda$CDM, the standard model of cosmology, have emerged from high-precision cosmic microwave background (CMB), baryon acoustic oscillations (BAO), and supernova (SN) observations. In this work, we focus on the two most significant challenges and argue that they provide two independent pieces of evidence pointing to the same underlying new physics beyond $\Lambda$CDM, sharpening arguments made recently in~\cite{Garny:2025kqj, Garny:2025szk}. 

The first challenge is the Hubble tension. The expansion rate of the Universe today measured directly by using standard candles disagrees with the expansion rate predicted in the $\Lambda$CDM model when fitted to CMB and BAO data. The most precise measurement of the expansion rate today from the SH0ES collaboration obtains $H_0 = 73.04 \pm1.04$ km/s/Mpc~\cite{Riess:2021jrx}, while the expansion rate inferred from the CMB, when assuming the $\Lambda$CDM model, implies $H_0 = 67.36 \pm 0.54$ km/s/Mpc \cite{Planck:2018vyg} when using Planck CMB data. These observational determinations of $H_0$ are in about $5\sigma$ conflict with each other. While one may wonder if there is unknown systematics affecting these measurements, which could explain this disagreement within the $\Lambda$CDM model, one should keep in mind that several measurements of the local expansion rate using different standard candles agree with the supernova measurement of SH0ES\footnote{For a recent review of local $H_0$ measurements, see \cite{H0DN:2025lyy}.}. At the same time, BAO measurements, combined with a constraint on the amount of baryons from Big Bang nucleosynthesis (BBN), agree with the CMB measurement \cite{DESI:2025zgx} only if $\Lambda$CDM is assumed. For this reason, we will focus on explanations of the Hubble tension, which involves new physics going beyond $\Lambda$CDM (for reviews of the tension and proposed solutions see~\cite{Schoneberg:2021qvd,Abdalla:2022yfr,CosmoVerseNetwork:2025alb}).

The second challenge, which we refer to as \textit{DESI anomaly}, is related to recent high-precision BAO measurement of the Dark Energy Spectroscopic Instrument (DESI). 
When DESI data is combined with Planck CMB \cite{Planck:2018vyg} and Pantheon+ supernova measurements~\cite{Scolnic:2021amr,Brout:2022vxf}, one finds a $2.8\sigma$ preference for a model with evolving dark energy parametrized as a two-parameter extension of $\Lambda$CDM. Specifically, if the dark energy equation-of-state parameter is assumed to take the form $w(a)=w_0 + w_a (1-a)$, where $\Lambda$CDM corresponds to $w_0=-1$, $w_a=0$, DESI finds~\cite{DESI:2025zgx} $w_0= - 0.838 \pm 0.055$, $w_a = - 0.62^{+0.22}_{-0.19}$. This interpretation is however unsettling as the preferred values would imply that dark energy entered a phantom regime with $w<-1$ in the recent past, although attempts to avoid this interpretation in terms of interacting dark energy and dark matter have been made~\cite{Khoury:2025txd,Kou:2025yfr,Chen:2025ywv,Liu:2025bss,Wang:2025znm,Caldwell:2025inn,SanchezLopez:2025uzw,Bedroya:2025fwh}. 

\begin{figure}[t]
  \centering
    \includegraphics[width=0.45\textwidth]{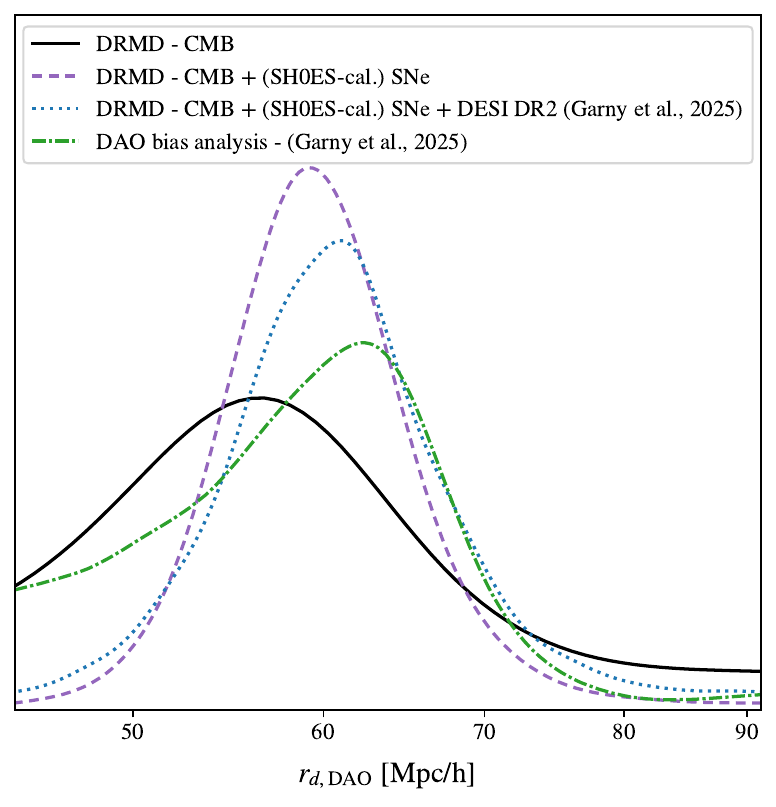}
    \vspace{-4mm}
  \caption{\footnotesize 1D posterior of the dark drag horizon, $r_{d,\mathrm{DAO}}$, from CMB alone (Planck 2018 temperature, polarization, and lensing) and from CMB combined with calibrated supernovae (Pantheon+ and SH$0$ES prior on the absolute magnitude $M$). Also shown are the corresponding posteriors from the four-parameter DRMD analysis of Ref.~\cite{Garny:2025kqj} (including DESI DR2) and the DAO bias study of Ref.~\cite{Garny:2025szk}. All curves are normalized to the same integral over the plotted interval. The compatibility of all curves with a value around $60\, \mathrm{Mpc}/h$ is one of the main results of this work. For reference, the baryon drag horizon is centred around $100\, \mathrm{Mpc}/h$.}
  \label{fig:r_DAO_1D}
\end{figure}

Another reason to find evolving dark energy unsettling as an explanation of the DESI anomaly is that evolving dark energy does not address the Hubble tension. The BAO experiments measure the product of the expansion rate $H_0$ and the sound horizon at baryon decoupling $r_{d,\mathrm{BAO}}$, which is measured by DESI  to be $H_0 \cdot r_{d,\mathrm{BAO}}   = 101.54 \,\pm \,0.73$ (100 km)/s~\cite{DESI:2025zgx}. This means, that in order to stay consistent with BAO measurements, any solution to the Hubble tension that fits the CMB with a higher value of $H_0$ needs to involve new physics before decoupling in order to also lower $r_{d,\mathrm{BAO}}$, and thus remain compatible with BAO measurements~\cite{Bernal:2016gxb,Knox:2019rjx}. This also implies that late-time evolving dark energy cannot resolve the Hubble tension.

One of the simplest ideas for new physics before recombination, which could lower the sound horizon and alleviate the Hubble tension, while staying compatible with BAO measurements, is dark radiation. Dark radiation would maintain a constant energy density relative to ordinary radiation in the radiation dominated epoch, and redshift away compared to the total energy density after matter-radiation equality shortly before recombination. This would lower the sound horizon and raise $H_0$ while minimizing its effect on structure formation in the late Universe. But if the dark radiation is free-streaming, it induces anisotropic stress and enhances diffusion damping on small scales in the CMB~\cite{Hou:2011ec} (predominantly scales that enter the horizon before matter-radiation equality) beyond an acceptable level. To avoid these problems, it is natural to consider a component of self-interacting dark radiation (SIDR), which behaves like a tightly coupled fluid, and is not free-streaming~\cite{Jeong:2013eza,Buen-Abad:2015ova,Buen-Abad:2017gxg,Archidiacono:2020yey,Blinov:2020hmc,Aloni:2021eaq}.

There are two basic problems with SIDR as a solution to the Hubble tension in its simplest form. First, it is ruled out by BBN~\cite{Schoneberg:2022grr}, which puts a rather stringent bound on any form of dark radiation. In order to alleviate the Hubble tension, an amount of SIDR, which corresponds to an extra amount of relativistic degrees of freedom $\Delta N_{\textrm{eff}} \gtrsim 0.5$, is needed, which is ruled out by BBN constraints. Second, even when ignoring BBN constraints, and allowing for $\Delta N_{\textrm{eff}} \gtrsim 0.5$, SIDR does not provide a full resolution of the Hubble tension, but only reduces the tension to around $3\sigma$ when using Planck 2018 CMB data~\cite{Garny:2024ums,Garny:2025kqj}. The latest constraints from ACT provide even stronger constraints on extra relativistic degrees of freedom \cite{AtacamaCosmologyTelescope:2025nti} (see section \ref{discussion} for a discussion of the latest ACT constraints). 

Fortunately, there is a simple and natural solution to the problems with SIDR as a solution to the Hubble tension. The first problem can be avoided if SIDR is created after BBN. This is exactly what happens in the Hot New Early Dark Energy (Hot NEDE) model\footnote{The general New Early Dark Energy (NEDE) framework for addressing the Hubble tension by a fast-triggered phase transition, and the general treatment of perturbations, with the possibility of the trigger field being either a scalar field, temperature, or something else, was first proposed in  \cite{Niedermann:2019olb, Niedermann:2020dwg}.} \cite{Niedermann:2021vgd,Niedermann:2021ijp,Garny:2024ums,Garny:2025kqj}, where SIDR is created from the latent heat of a supercooled phase transition in the dark sector after BBN but before recombination \cite{Garny:2024ums,Garny:2025kqj}. The second problem can be avoided if SIDR interacts with a fraction of dark matter early on, but decouples from dark matter just around equality in a process we refer to as dark radiation-matter decoupling (DRMD) \cite{Garny:2025kqj}. In Hot NEDE, this is naturally achieved in a model where the phase transition is related to the spontaneous symmetry breaking of $SU(N)\to SU(N-1)$~\cite{Garny:2025kqj}. 
In this setting, it has been shown that the Hubble tension can be reduced below $1.5\sigma$ when using Planck 2018 CMB, DESI DR2 BAO and Pantheon+ supernova data \cite{Garny:2025kqj}.

Other scenarios exhibit a similar decoupling feature. In the \textit{Atomic Dark Matter} model, dark atoms form within the dark sector in a process analogous to recombination in the visible sector~\cite{Kaplan:2009de,Cyr-Racine:2012tfp,Cyr-Racine:2021oal,Blinov:2021mdk,Bansal:2022qbi}. In the \textit{Stepped Partially Acoustic Dark Matter} model, decoupling instead occurs when a massive component of the dark radiation fluid annihilates away~\cite{Buen-Abad:2022kgf}. These scenarios have been explored as possible resolutions of the Hubble tension~\cite{Bansal:2021dfh,Buen-Abad:2023uva,Buen-Abad:2024tlb}, and some achieve a comparable level of phenomenological success for the datasets mentioned above~\cite{Buen-Abad:2025bgd}. In their present form, they lack, however, a mechanism for creating dark radiation after BBN.

In the case of Hot NEDE, the decoupling mechanism was tested against data using an effective fluid description, which we refer to as the DRMD model~\cite{Garny:2025kqj}. It extends $\Lambda$CDM by three additional parameters: the amount of (self-interacting) dark radiation produced after BBN, parametrized by $\Delta N_{\textrm{eff}}$; the fraction of interacting dark matter, $f_\textrm{idm}$; and the redshift of dark decoupling, $z_\textrm{dec}$. Both the previous results and those presented in this work therefore apply more broadly to scenarios in which a dark radiation component interacts with a fraction of dark matter and subsequently decouples around matter–radiation equality via an exponential suppression of the drag rate.

Just as decoupling creates a BAO feature in the matter power spectrum at a characteristic scale given by the sound-horizon when baryons decouple from the radiation bath of photons, in a similar way the dark matter and dark radiation interactions will, together with gravity, create acoustic oscillations in the dark fluid, and the DRMD process will create a Dark Acoustic Oscillation (DAO) at a characteristic scale set by the dark sound horizon at the time of dark decoupling~\cite{Cyr-Racine:2012tfp,Cyr-Racine:2013fsa,Garny:2025kqj,Garny:2025szk}. Thus the existence of a DAO at a characteristic scale, as well as its amplitude, is a prediction of this class of models and closely tied to their ability to resolve the Hubble tension. The aim of this work is to sharpen and quantify this connection between DAO and the Hubble tension.

In previous work~\cite{Garny:2025szk}, we showed that the existence of a DAO close in scale to the actual BAO, could bias the extraction of the BAO scale in the DESI analysis, if ignored. We furthermore showed that accounting for the DAO bias can explain the DESI DR2 anomaly as an alternative explanation to evolving dark energy. We also found that the DAO scale and amplitude, preferred by DESI compared to no DAO, matches the properties predicted by Hot NEDE with DRMD when resolving the Hubble tension, providing a first validation of DRMD as a solution to the Hubble tension \cite{Garny:2025kqj}.

The DAO predicted by DRMD can be further searched for in DESI full-shape data and future large-scale-structure surveys, such as Euclid \cite{Euclid:2024yrr}, or the Roman Space Telescope \cite{Spergel:2015sza}. To facilitate searches for the DAO predicted by DRMD-like solutions to the Hubble tension, we will here study the DAO predictions made by this class of solutions to the Hubble tension independently of large-scale structure observations like DESI. For this purpose we will extract the preferred scale and amplitude of the DAO when fitting the DRMD model to CMB and supernova data only. This prediction provides a target for current and future large-scale structure surveys. In particular we show that these predictions are consistent with results already extracted from the DESI DR2 data in \cite{Garny:2025kqj}, which, as discussed above, provides a preliminary validation of the importance of DRMD in resolving the Hubble tension. 

The posterior distributions shown in Fig.~\ref{fig:r_DAO_1D} summarize the main results of this work. Starting from CMB data alone (black), we then add SH$0$ES-calibrated supernova data (purple dashed), and finally compare with a previous analysis in~\cite{Garny:2025kqj} that also included BAO data (blue dotted). The resulting constraints remain mutually consistent throughout.
Moreover, they are also consistent with the DAO scale extracted independently of the DRMD model from DESI DR2 data, when taking into account how a DAO peak of a given scale would bias the extraction of the BAO scale~\cite{Garny:2025szk}. 

This paper is organized as follows. In section \ref{model}, we review the effective fluid model of DRMD, discuss why a decoupling feature is needed for solving the Hubble tension within this class of models, and explain the origin of the DAO feature in the DRMD model. In  section \ref{results} we discuss the results from our parameter inference, which includes targets for future full-shape analyses of the DESI data and future large-scale structure telescopes on ground or in space. We conclude with a discussion and outlook in section \ref{discussion}.

\section{Dark Radiation-Matter Decoupling}\label{model}

The decoupling between dark radiation and dark matter within Hot NEDE is described by an effective fluid implementation, which we refer to as DRMD and which has also been implemented into the publicly available Boltzmann code \texttt{DRMD-CLASS}\footnote{\href{https://github.com/NEDE-Cosmo/DRMD-CLASS.git}{https://github.com/NEDE-Cosmo/DRMD-CLASS.git}}~\cite{Garny:2025kqj}. The DRMD mechanism and its implementation is however sufficiently general that the code can also be used to describe the phenomenology of certain other proposed scenarios falling within this class of models \cite{Buen-Abad:2022kgf, Buen-Abad:2023uva}. For concreteness, we first briefly review the origin of a decoupling between dark radiation and dark matter in Hot NEDE, and comment on how our results apply more generally to other scenarios. After that, we proceed to describing the DRMD model in Sec.~\ref{DRMDEFM}, before discussing the resolution of the Hubble tension in Sec.~\ref{sec:H0} and DAO in Sec.~\ref{sec:DAO}.

In the Hot NEDE model of \cite{Garny:2024ums,Garny:2025kqj}, 
the dark sector is described by a weakly coupled $SU(N)$ gauge theory, including a light dark Higgs field $\Psi$ and a  heavy Dirac  fermion dark matter multiplet $\chi$, both of which transform under the fundamental representation of $SU(N)$, but are not coupled to the Standard Model sector. The Higgs field leads to spontaneous symmetry breaking from $SU(N)$ to $SU(N-1)$ via a supercooled phase transition, taking place in the epoch after BBN but well before recombination.
The associated latent heat creates a radiation plasma consisting of massless gauge bosons, $A^a$, of the unbroken $SU(N-1)$ and the physical dark Higgs boson $\psi$, which arises when the complex scalar $\Psi$  picks up a non-vanishing vacuum  expectation value $v$, {\it i.e.}\ $|\Psi| \equiv (v + \psi)/\sqrt{2}$. The gauge interactions between the relativistic degrees of freedom lead to a tightly coupled fluid, which behaves like SIDR. Such a scenario is naturally realized for a Coleman-Weinberg potential with small explicit breaking of classical conformal invariance~\cite{Coleman:1973jx,Witten:1980ez}. This dark sector re-heating raises the effective number of relativistic degrees of freedom to\footnote{There is an additional small change to $\Delta N_\text{eff}$ when the Higgs boson becomes non-relativistic, but that is a sub-leading effect, which  vanishes in the large-$N$ limit, and so we ignore it here for simplicity. For a detailed discussion, see \cite{Garny:2024ums}.}
\begin{equation}\label{eq:N_eff}
 \Delta N_\textrm{eff} \equiv \Delta N_\textrm{NEDE}\sim \left(\frac{g\,v}{85\,\text{keV}}\right)^4\left(\frac{10^8}{1+z_*}\right)^4 \quad ~ (\text{for}~~N=3)~, 
\end{equation}
where $g$ is the $SU(N)$ gauge coupling and $z_*$ denotes the redshift of the phase transition that must occur after BBN but before the first scales probed through the CMB start entering the horizon, implying $10^6 \lesssim z_* \lesssim   10^9$. 

Before the phase transition a fraction $f_{\textrm{idm}}$ of dark matter is in a multiplet, $\chi$, charged under $SU(N)$ in the fundamental representation.  After the phase transition, $N-1$ components of the dark matter multiplet, $\chi_j$, are charged under the unbroken $SU(N-1)$ and interact with the dark radiation fluid, while the component in the broken direction $\chi_0$ is neutral under the unbroken $SU(N-1)$. Thus, $\chi_0$ only interacts with the $\chi_j$ through dark matter scattering mediated by the massive gauge bosons with components along the broken directions $SU(N)/SU(N-1)$,
maintaining equilibrium between $\chi_j$ and $\chi_0$. Accordingly, after the phase transition, we can separate the fraction of interacting dark matter, $f_{\textrm{idm}}$, into a neutral, $f_\mathrm{\chi_0}(z)$, and charged, $f_\mathrm{\chi_j}(z)$, fraction: $f_{\textrm{idm}}=f_\mathrm{\chi_j}(z)+f_\mathrm{\chi_0}(z)$.

Initially after the phase transition, the fraction of interacting dark matter, $f_{\textrm{idm}}$, is in equilibrium with the dark radiation fluid  through $t$-channel Compton scattering between the massless $SU(N-1)$ gauge bosons, $A^a$, and the charged dark matter component $\chi_j$, while the $\chi_0$ and $\chi_j$, as mentioned above, maintain equilibrium with each other through $t$-channel dark matter scattering mediated by the massive gauge bosons along the broken directions. As a result, initially the $\chi$ particles and massless gauge bosons form a  tightly coupled fluid similar to the photon-baryon fluid in the visible sector (for more details, see~\cite{Garny:2025kqj}).

Due to the fact that the $\chi_j$ and the $\chi_0$ do not both couple to the massless gauge bosons $A^a$, their masses will receive different one-loop corrections after the symmetry breaking, leading to a small mass gap $\Delta m \approx g^3 v/(32 \pi)$. Crucially, the neutral component $\chi_0$ is lighter than its charged counterpart $\chi_j$. This mass splitting is an unavoidable quantum effect. Therefore, as the temperature in the dark fluid, $T_d$, drops below the mass gap $\Delta m$, the charged dark matter component $\chi_j$ will become Boltzmann suppressed and convert into the neutral component $\chi_0$. The fraction $f_\mathrm{\chi_j}(z)$ relative to $f_{\textrm{idm}}$ can be therefore be written as
\begin{align}
\left(\frac{f_\mathrm{\chi_j}(z)}{f_{\textrm{idm}}}\right)_\text{eq}  = \frac{(N-1)\mathrm{e}^{-\Delta m / T_d } }{(N-1)\mathrm{e}^{-\Delta m / T_d } + 1}\,.
\end{align}
Right after the phase transition, when $T_d \gg \Delta m$ the fractions satisfy equipartition
\begin{align}
\left(\frac{f_\mathrm{\chi_j}(z)}{f_{\textrm{idm}}}\right)_\text{eq}   \to \frac{(N-1)}{N}\,,
\end{align}
but as the temperature later drops below the mass gap, $T_d < \Delta m$, the fraction of charged dark matter gets exponentially suppressed
\begin{align}
\left(\frac{f_\mathrm{\chi_j}(z)}{f_{\textrm{idm}}}\right)_\text{eq}  \to (N-1)\,\mathrm{e}^{-\Delta m / T_d }\,.
\end{align}
We denote the redshift at which the Boltzmann suppression becomes active at $T_d \approx \Delta m$ by $z_{\textrm{stop}}$. In other words, we parametrize
\begin{align}\label{eq:z_stop}
\exp{\left(-\frac{\Delta m}{  T_d} \right)}=\exp \left( -\frac{1+z_\mathrm{stop}}{1+z} \right)\,.
\end{align}
Tight coupling between dark matter and dark radiation is maintained as long as the associated drag rate from $t$-channel Compton scattering within the $SU(N−1)$ gauge sector satisfies $\Gamma_\textrm{drag} \gg H$. The drag rate is given by
\begin{align}
\Gamma_\mathrm{drag} = \frac{f_{\chi_j}(z)}{f_{\textrm{idm}}} \Gamma_{\chi_jA_\mu^a}\,,
\end{align}
where $\Gamma_{\chi_jA_\mu^a}$ is the $t$-channel Compton interaction rate of $\chi_j$ states with the radiation bath of $A^a$ bosons \cite{Garny:2025kqj}. The ratio $\Gamma_{\chi_jA_\mu^a}/H$ is constant during radiation domination~\cite{Buen-Abad:2015ova,Rubira:2022xhb}, however, we infer that $\Gamma_\mathrm{drag}/H$ rapidly decreases once Boltzmann suppression kicks in for $z<z_\textrm{stop}$, and once the ratio drops below unity the dark radiation cannot maintain tight coupling to the non-relativistic $\chi$-components,
leading to the decoupling between dark matter and dark radiation. We denote the corresponding redshift when $\Gamma_\mathrm{drag}= H$ by $z_\textrm{dec}$.

Before proceeding to discuss the effective fluid model of DRMD, let us pause to emphasize that the above discussion applies to any model or scenario where a fraction of dark matter, 
$f_{\textrm{idm}}$, transitions from being tightly coupled to self-interacting dark radiation to exponentially decoupling over the redshift interval $z_\textrm{stop} > z > z_\textrm{dec}$.
Consequently, the DRMD fluid model described below, as well as the associated phenomenological results, are applicable to any scenario exhibiting this particular decoupling behavior, see {\it e.g.}~\cite{Kaplan:2009de,Cyr-Racine:2012tfp,Cyr-Racine:2021oal,Blinov:2021mdk,Bansal:2022qbi,Bansal:2021dfh,Buen-Abad:2022kgf,Schoneberg:2023rnx,Buen-Abad:2023uva,Buen-Abad:2024tlb,Buen-Abad:2025bgd}.

\subsection{Effective Fluid Model of Dark Radiation-Matter Decoupling}\label{DRMDEFM}

At the fluid level, the perturbations in the dark sector are described  by adding two new components, dark radiation and interacting dark matter ($\chi$), to the standard $\Lambda$CDM model. As discussed above, due to the non-Abelian self-interactions of gauge bosons,
the cosmological dynamics of the relativistic degrees of freedom can be described as a single fluid in terms of a density contrast $\delta_{\textrm{DR}}$ and a velocity divergence $\theta_\textrm{DR}$. The initial energy density of the dark radiation fluid $\rho_\mathrm{DR}$ is parametrized in terms of its contribution to the effective number of neutrino species, $\Delta N_\mathrm{eff}$. Concretely, we define
\begin{align}\label{eq:N_eff_2}
\left(\frac{\rho_\mathrm{DR}}{\rho_{1\nu}} \right)_\mathrm{ini}
= \Delta N_\mathrm{eff}\,,
\end{align}
where $\rho_{1\nu} = \frac{7}{4}\frac{\pi^2}{30}(T_{\nu,0}/a)^{4}$ denotes the energy density of a single massless neutrino species with temperature $T_{\nu,0}$.
In the context of Hot NEDE this `initial' value is understood as the value {\it after} the supercooled phase transition that heats the dark sector, {\it i.e.} after BBN but before the modes observed in the CMB enter the horizon.

The non-relativistic, interacting dark matter
component, with the neutral $\chi_0$ and charged $\chi_j$ states assumed to be tightly coupled to each other, can be described by a pressureless fluid with density contrast $\delta_\chi$
and velocity divergence $\theta_\chi$. The corresponding energy density $\rho_\chi$ is measured relative to the total dark matter density
\begin{align}\label{eq:f_idm_2}
\rho_\mathrm{\chi}=f_\mathrm{idm} (\rho_\mathrm{cdm }+\rho_\mathrm{\chi})\,,
\end{align}
where  $\rho_\mathrm{cdm }$ is the cold dark matter density.
In synchronous gauge, the perturbation equations of the interacting dark matter component are
\begin{subequations}\label{eq:pert_chi}
\begin{align}
\delta_\chi^\prime + \left( \theta_\chi + \frac{h^\prime}{2} \right) &= 0 ~,  \label{eq:delta_idm}\\
\theta_\chi^\prime + \mathcal{H} \theta_\chi &= \mathcal{G} \Delta^\prime ~, \label{eq:theta_idm}
\end{align}
\end{subequations}
where $h$ is the trace of the spatial components of the metric fluctuation, and we introduced the velocity slip defined by $\Delta'=\theta_\mathrm{DR}-\theta_\chi$.
The time-dependent function $\mathcal{G}$ quantifies the momentum-transfer between (interacting) dark matter and dark radiation and is determined by the drag-rate $\Gamma_\textrm{drag}$. Its time dependence can be parametrized through \cite{Garny:2025kqj}
\begin{align} \label{eq:G}
\mathcal{G} \simeq \mathcal{H} 
\left(\frac{\mathcal{G}}{\mathcal{H}}\right)_\mathrm{ini} \left( 1+ \frac{\Omega_m}{\Omega_\mathrm{rad} } \frac{1}{1+z} \right)^{-1/2} \, \exp \left( -\frac{1+z_\mathrm{stop}}{1+z} \right)~,
\end{align}
where for Hot NEDE the redshift-independent part can be related to the microphysical $t$-channel Compton interaction rate of the charged dark matter states ($\chi_j$) with the radiation bath of the massless gauge bosons ($A_\mu^a$)
\begin{equation}
 \left[\frac{\mathcal{G}}{\mathcal{H}}\right]_\mathrm{ini} 
 = (N-1) \frac{a^2  \Gamma_{\chi_jA_\mu^a}}{a\mathcal{H}}~,
\end{equation}
where the right hand side is constant during radiation domination (after the Higgs has become non-relativistic and no longer contributes to the dark radiation fluid). 
The decoupling redshift $z_\mathrm{dec}$ is then defined through
\begin{align}\label{eq:z_dec}
\mathcal{G}(z_\mathrm{dec}) = 
\mathcal{H}(z_\mathrm{dec})~.
\end{align}
Together with \eqref{eq:G}, it implies
\begin{align}\label{eq:G_ini}
\frac{1+z_\textrm{stop}}{1+z_\textrm{dec}} & 
\simeq \ln\left( \left[\frac{\mathcal{G}}{\mathcal{H}}\right]_\mathrm{ini} \right) ~.
\end{align}

Similarly to the perturbation equations for the interacting dark matter component, the perturbation equations of the dark radiation fluid are \cite{Garny:2025kqj}
\begin{subequations}\label{eq:pert_DR}
\begin{align}
\delta_\mathrm{DR}^\prime 
+ \left( 1+ w_{\rm DR} \right)\left(\theta_{\rm DR} + \frac{h^\prime}{2}\right)  + 3 \mathcal{H} \left(c_s^2 - w_{\textrm{DR}} \right) \delta_\mathrm{DR}&= 0~,  \label{eq:delta_DR}\\
\theta^\prime_\mathrm{DR} - \frac{k^2 c_s^2}{1 + w_{\rm DR}} \delta_{\rm DR} + \mathcal{H} \left(1-3 c_s^2  \right) \theta_\mathrm{DR} &= -\mathcal{G}R \Delta^\prime   \label{eq:theta_DR}~,
\end{align}
\end{subequations}
with $R$ follows from ensuring conservation of the combined dark matter-radiation fluid,
\begin{align}\label{eq:R}
R = \frac{1}{1+w_\textrm{DR}} \frac{\rho_\chi}{\rho_\mathrm{DR}}~,
\end{align}
and the sound speed in the radiation fluid
\begin{align}\label{eq:c_s}
c_s^2= w_\mathrm{DR} - \frac{w_\textrm{DR}^\prime}{3 \mathcal{H} (1+w_\mathrm{DR})}~.
\end{align}
We keep the equations general enough to allow for a time-dependence of the equation-of-state parameter $w_\mathrm{DR}$.
Since all the relevant scales are initially super-horizon, we can impose adiabatic super-horizon initial conditions (assuming that early enough $w_\mathrm{DR} = 1/3$)~\cite{Garny:2025kqj} 
\begin{align}
\frac{4}{3}\delta_\chi = 
\delta_\textrm{DR}= \delta_\mathrm{\gamma} = \delta_\mathrm{\nu}= -\frac{2}{3} C k^2 \tau^2~,
\end{align}
where  $C$ is an integration constant determined by matching to the co-moving curvature perturbation, while $\delta_\gamma$ is the photon and $\delta_\nu$ the neutrino density contrast. At leading non-trivial order in $k\tau$, the perturbation equations imply the
initial conditions for the dark radiation perturbations
\begin{align}\label{eq:ini_pert_1}
\theta_\textrm{DR} = \theta_\mathrm{\gamma} = - \frac{1}{18} C k^4 \tau^3 ~,
\end{align}
and the dark matter velocity perturbations
\begin{align}\label{eq:ini_pert_2}
\theta_\chi = \frac{(\mathcal{G}/\mathcal{H})_\textrm{ini}}{4 + (\mathcal{G}/\mathcal{H})_\mathrm{ini}} \theta_\mathrm{\gamma}~.
\end{align}
\texttt{DRMD-CLASS} uses the dynamical equations in \eqref{eq:pert_chi} and \eqref{eq:pert_DR} supplemented by the initial conditions in \eqref{eq:ini_pert_1} and \eqref{eq:ini_pert_2}. The redshift-dependence of the momentum transfer function $\mathcal{G}$ is parametrized through \eqref{eq:G}. At the same time, $z_\mathrm{dec}$ as defined in \eqref{eq:z_dec} is inferred from the background evolution through linear interpolation on the integration grid.
\subsection{The Hubble Tension}\label{sec:H0}

The angular size $\theta_s$ of the sound horizon  in the CMB, which is very precisely determined by the position of the first peak, is given by the ratio of the comoving sound horizon $r_s(z_{LS})$ and the angular diameter distance $d_a(z_{LS})$ evaluated at the last scattering surface at $z=z_{LS}$.
Since the Universe is matter dominated at the time of ordinary baryon-photon decoupling, the angular diameter distance is approximately determined by the inverse Hubble rate today and $\Omega_m$. Thus, one finds 
\begin{align}
\theta_s^{-1} = \frac{d_a(z_{LS})}{r_s(z_{LS})} \simeq \frac{1}{H_0r_s} \int_0^{z_{LS}} dz\, \left[\Omega_m (1+z)^3+(1-\Omega_m)\right]^{-1/2}\,.
\end{align}
As a result once $r_{d,\mathrm{BAO}} \cdot H_0 \simeq r_s \cdot H_0$ is fixed by BAO measurements, $\Omega_m$ is determined by the relation between the angular size of the sound horizon measured in CMB and $r_{d,\mathrm{BAO}} \cdot H_0$ measured by the BAO position independently of the precise value of $H_0$ itself.

However, since the relative matter density parameter $\Omega_m$ is fixed independent of $H_0$, the physical matter density $\omega_m = \Omega_m h^2$, depends sensitively on the value of $H_0 \equiv h \,100\,\mathrm{km/s/Mpc}$. In particular, resolving the $H_0$ tension requires a larger than $10\%$ increase in the physical matter density $\omega_m$. Since the small-scale power in the CMB and the matter power spectrum depends on $\omega_m$, models solving the Hubble tension tend to predict more power on small scales than predicted within $\Lambda$CDM. This effect is particularly severe for modes that enter \textit{before} radiation-matter equality and are thus exposed to radiation-driving effects, which are sensitive to $\omega_m$.  

\begin{figure}[th]
  \centering
    \includegraphics[width=0.94\textwidth]{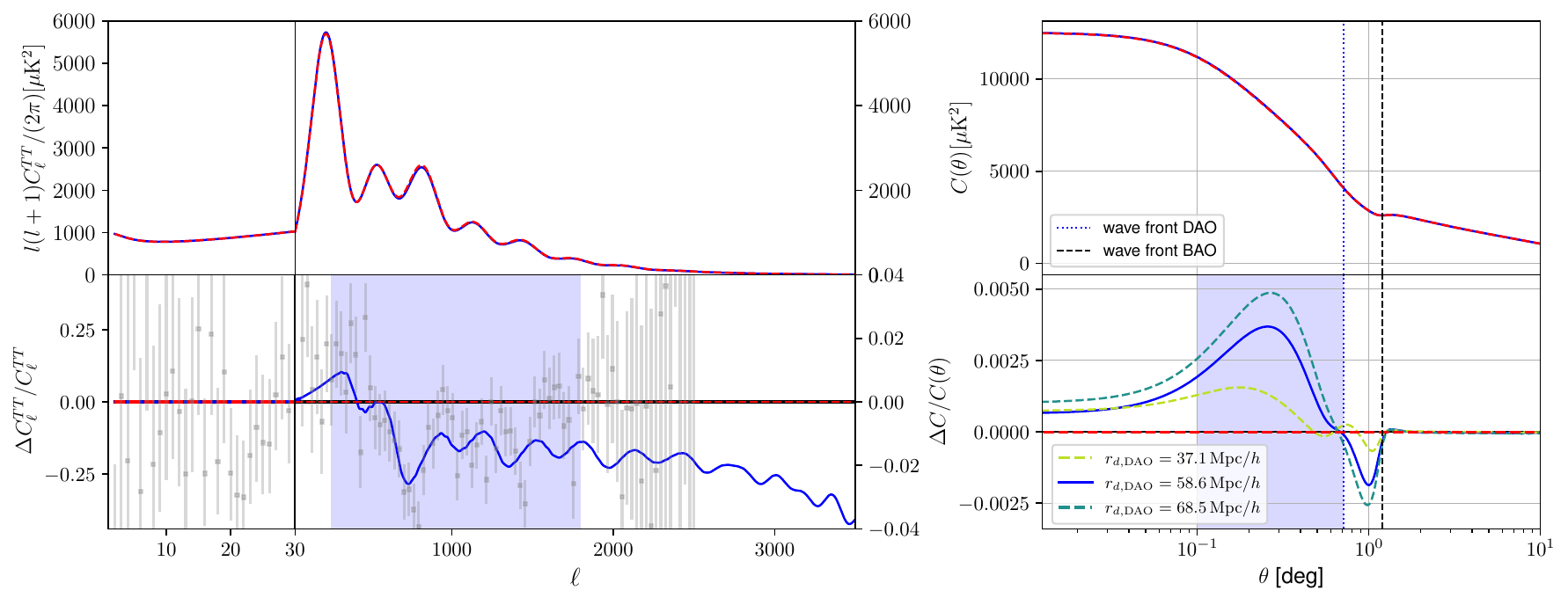}
    \vspace{-4mm}
\includegraphics[width=0.55\textwidth]{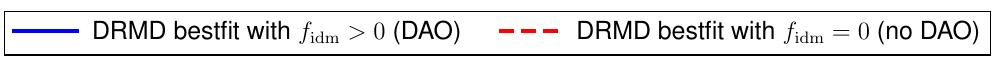}
\vspace{-1mm}
\caption{\footnotesize
CMB temperature anisotropies for the global best-fit obtained from our combined analysis with CMB and SH$0$ES-calibrated supernova data. We compare DRMD with DAO ($f_\mathrm{idm}>0$, blue) and without DAO ($f_\mathrm{idm}=0$, red dashed). The lower panels show the relative difference.
\textit{Left:} In harmonic space, the DAO imprint percent-level oscillatory features in the temperature power spectrum $C_\ell^\mathrm{TT}$, together with a broadband suppression induced by enhanced decay of the gravitational potential. Planck residuals are shown as gray error bars.
We note the change in spacing of the axes as separated by the vertical line at $\ell=30$. \textit{Right:} In position space, this imprint appears as a localized feature in the angular correlation function $C(\theta)$ (blue-shaded band) at the permille level. The coloured dashed lines correspond to different choices of the DAO drag horizon scale.
}
  \label{fig:CMB}
\end{figure}

The DRMD mechanism is crucial for solving the Hubble tension, while not encountering this problem with too much power on small scales. In contrast, SIDR models (without an early coupling between dark matter and dark radiation) end up being limited in their ability to solve the Hubble tension by predicting too much power on small scales. A component of interacting dark matter will, due to its interactions with dark radiation, clump less than cold dark matter. It is dragged by dark radiation and effectively experiences a small positive pressure, suppressing the gravitational potential for any mode entering the horizon before it decouples at $z_\mathrm{dec}$. For it to compensate the effect of a higher $\omega_m$  on  only small scales entering before matter domination, but leave the large scales invariant, the interacting dark matter needs to decouple from dark radiation at around matter domination, {\it i.e.}\ $z_\mathrm{dec} \sim z_\mathrm{eq}$~\cite{Garny:2025kqj}. This small-scale suppression  is shown in the left panel of Fig.~\ref{fig:CMB}, which compares a typical DRMD temperature power spectrum (blue) to the corresponding SIDR spectrum with $f_\mathrm{idm}=0$ (red dashed).
When this happens, DRMD allows for a full resolution of the Hubble tension when fitted to Planck 2018 CMB (and other data sets such as DESI DR2 BAO, and Pantheon+ supernova data~\cite{Garny:2025kqj}; see also left panel in Fig.~\ref{fig:2D_posteriors}). New data from ground based CMB experiments, such as ACT \cite{AtacamaCosmologyTelescope:2025nti} and SPT \cite{SPT-3G:2025bzu}, provide additional constraints on small scales. As we will discuss below, the imprint of the DAO in the CMB is however localized on intermediate scales best probed by Planck, and is therefore a robust prediction of the DRMD model's ability to address the Hubble tension (see also section \ref{discussion}).

\subsection{Dark Acoustic Oscillations}\label{sec:DAO}

Initially dark matter and dark radiation are tightly coupled. In this regime the two fluids share a common velocity divergence,
\begin{align}
\theta_\mathrm{DR} = \theta_\chi \equiv \theta\,,
\end{align}
which we denote as $\theta$. Using Eqs.~\eqref{eq:delta_idm} and \eqref{eq:delta_DR}, and assuming adiabatic initial conditions at early times where $w_\mathrm{DR} = 1/3$, we obtain
\begin{align}
\frac{\delta_\mathrm{DR}}{1+w_\textrm{DR}} = \delta_\chi~.
\end{align}
Furthermore, combining Eqs.~\eqref{eq:theta_idm} and \eqref{eq:theta_DR} yields
\begin{align}
\mathcal{G}\Delta' 
= 3\, c_{s,\mathrm{eff}}^2 
\left[\frac{k^2}{4} \delta_\mathrm{DR} + \mathcal{H} \theta \right] ,
\end{align}
where we have defined
\begin{align} \label{eq:sound_speed}
c_{s,\textrm{eff}}^2 = \frac{c_s^2}{1+R}~.
\end{align}

It is convenient to introduce the total density perturbation
\begin{align}
\delta \rho = \delta \rho_\chi + \delta \rho_\mathrm{DR}\,,
\qquad
\rho = \rho_\chi + \rho_\mathrm{DR}\,,
\end{align}
and to rewrite the system in terms of the corresponding combined fluid variables $\delta= \delta \rho/\rho$ and $\theta$. After straightforward algebra, the perturbation equations reduce to two independent equations,
\begin{subequations}
\begin{align}
\delta' 
+ \left(1 + w_{\rm eff}\right)\left(\theta + \frac{h'}{2}\right) 
+ 3 \mathcal{H} \left(c_{s,\mathrm{eff}}^2 - w_{\rm eff} \right)\delta
&= 0\,, \label{eq:delta_DR_combined} \\
\theta' 
- \frac{k^2 c_{s,\mathrm{eff}}^2}{1 + w_{\rm eff}} \delta 
+ \mathcal{H} \left(1 - 3 c_{s,\mathrm{eff}}^2 \right)\theta 
&= 0\,.
\end{align}
\end{subequations}
These are precisely the perturbation equations of a single fluid~\cite{Ma:1995ey} with time-dependent equation-of-state parameter
\begin{align}
w_{\rm eff} = \frac{p_\chi + p_\mathrm{DR}}{\rho_\chi + \rho_\mathrm{DR}}= \frac{w_\mathrm{DR}}{1+(1+w_\mathrm{DR})R}\,.
\end{align}
In particular, using the definition in  \eqref{eq:sound_speed} alongside \eqref{eq:R} and \eqref{eq:c_s}, we can indeed verify that $c_{s,\mathrm{eff}}$ as defined in \eqref{eq:sound_speed} agrees with the expression for the sound speed of a barotropic fluid
\begin{align}
 \frac{p'}{\rho'}=w_\mathrm{eff}-\frac{w_\mathrm{eff}'}{1+3\mathcal{H}(1+w_\mathrm{eff})}=c_{s,\mathrm{eff}}^2\,.
\end{align}
This in turn allows us to define the dark acoustic (or dark drag) horizon,
\begin{align}\label{eq:r_D}
r_{d,\mathrm{DAO}} 
= \int_0^{\eta_{\mathrm{dec}}} \mathrm{d}\eta \, c_{s,\mathrm{eff}}(\eta)
= \int_{z_\mathrm{dec}}^\infty \mathrm{d}z\, 
\frac{c_{s,\mathrm{eff}}(z)}{H(z)}\,,
\end{align}
which corresponds to the comoving distance a dark acoustic wave can travel prior to decoupling. As we will see below, this scale sets the characteristic angular size of DAO features in the CMB and in the matter distribution of the Universe.

\begin{figure}[t]
  \centering
    \includegraphics[width=0.94\textwidth]{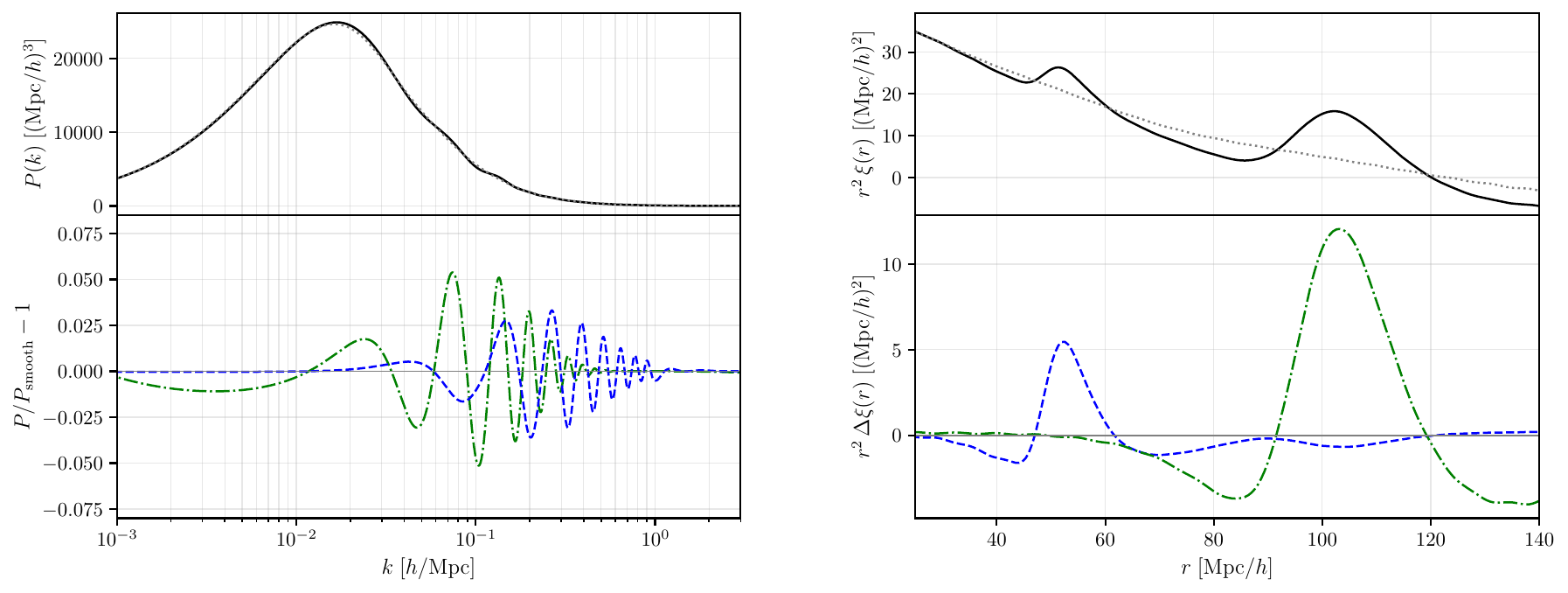}
        \vspace{-2mm}
\includegraphics[width=0.50\textwidth]{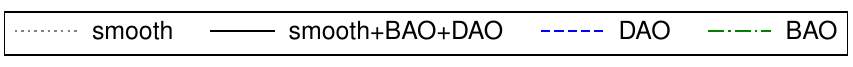}
\vspace{-4mm}
  \caption{\footnotesize BAO-DAO decomposition of the linear matter power spectrum $P_\text{lin}(k)$ (left) and corresponding correlation function $\xi(r)$ (right) for the global best-fit obtained from our combined analysis with CMB and SH$0$ES-calibrated supernova data. 
  The lower panels show the DAO (blue) and BAO (green) features relative to the smooth broadband contribution (dotted lines in the upper panels). For details on the decomposition into smooth and wiggly BAO as well as DAO contributions, see section  \ref{results}.
  }
  \label{fig:DAO_BAO_split}
\end{figure}

To obtain an initial estimate, we note that the DRMD process is analogous to standard photon decoupling in the visible sector, although it occurs somewhat earlier, close to matter–radiation equality. In $\Lambda$CDM, the acoustic scale associated with baryon decoupling is
$k_{A,\mathrm{dec}} = {\pi}/{r_{d,\mathrm{BAO}}} \approx 0.022 \ \mathrm{Mpc}^{-1}$,
while the scale associated with matter–radiation equality is
$
k_{A,\mathrm{eq}} = {\pi}/{r_{d,\mathrm{eq}}} \approx 0.050 \ \mathrm{Mpc}^{-1}, 
$ where the subscript $A$ denotes the acoustic scale, which is half the full angular scale. The ratio of these two characteristic scales is therefore
$k_{A,\mathrm{dec}}/k_{A,\mathrm{eq}}\approx 0.44$.
Similar to the standard BAO feature at $r_{d,\mathrm{BAO}} \approx 100 \, \mathrm{Mpc}/h$ (see Tab.~\ref{tab:MCMC}), one would then naively expect a DAO feature at roughly half that distance
$r_{d,\mathrm{DAO}} \approx 50 \, \mathrm{Mpc}/h$.
In Fig.~\ref{fig:DAO_BAO_split}, we extract the BAO and DAO contributions from the best-fit DRMD model to the linear matter power spectrum $P_\text{lin}(k)$ as well as the corresponding correlation function $\xi(r)$, and indeed find the DAO feature located close to $r_{d,\mathrm{DAO}} \approx 50 \, \mathrm{Mpc}/h$, in agreement with this simple estimate.

Before returning to the properties of these DAO features and their impact on matter clustering in more detail, let us stress that their effect on the CMB angular power spectrum is much smaller. This is because the DM-DR interaction directly influences the matter perturbations, while the impact on photons is only indirect via the associated metric perturbations. More precisely, the relative size of  DAO versus BAO features in the matter distribution is controlled by the ratio of densities of interacting dark matter and baryons, {\it i.e.} $f_\text{idm}/f_b$ where $f_b$ is the baryon fraction. On the other hand, for the CMB, the acoustic peaks driven by sound waves of the photon-baryon plasma are an ${\cal O}(1)$ effect, while the impact of DAO is transmitted only via their impact on gravitational potentials, which is suppressed as $(\rho_\text{idm}+\rho_\text{DR})/\rho_\text{tot}$. This ratio is of order $0.13\, \Delta N_\mathrm{eff}$ before equality and goes to $f_\text{idm}$ thereafter.

For the CMB (neglecting the difference between the the sound horizon at baryon decoupling and at last scattering) these scales subtend angles
$\theta_{A,\mathrm{dec}} = {r_{d,\mathrm{BAO}}} / {d_A(z_{LS})} \approx 0.01 \, \mathrm{rad} \approx 0.6^\circ $,
$\theta_{\mathrm{eq}} = {r_{d,\mathrm{eq}}}/ {d_A(z_{LS})} \approx 0.005 \, \mathrm{rad} \approx 0.3^\circ $,
where we have used $d_A(z_{\mathrm{LS}}) = 13.9 \, \mathrm{Gpc}$ for the angular diameter distance. 
These angular scales can be identified directly in position space via the angular correlation function
\begin{equation}
C(\theta)
=
\sum_{\ell}
\frac{2\ell+1}{4\pi}\,
C_\ell\,
P_\ell(\cos\theta)\,,
\end{equation}
shown in the right panels of Fig.~\ref{fig:CMB}. In particular, the estimate $\theta_{\mathrm{eq}} \approx 0.3^\circ$ corresponds to the bump visible in the lower-right panel of Fig.~\ref{fig:CMB}, where the blue line denotes the DRMD best-fit cosmology. This feature extends over a finite range of angular scales (indicated by the blue-shaded region) that is especially well probed by Planck 2018 (see residuals in lower left panel). For comparison, we also display models with smaller and larger drag horizons (dashed lines), which shift the feature to correspondingly smaller and larger angles.

The small dip at slightly larger scales arises from the broadband modification induced by the dark matter–dark radiation interaction, which affects the BAO. The acoustic interpretation of both the peak and dip becomes even clearer when displaying the wave fronts, which represent the causal propagation limit and are expected at twice the corresponding horizon scale~\cite{Bashinsky:2000uh}. In our case, they occur at angular scales
 $2\,r_{d,\mathrm{DAO}}/d_A{(z_{LS})}\approx 0.6^\circ$ (blue dotted) for the DAO and at $2\,r_{d,\mathrm{BAO}}/d_A{(z_{LS})}\approx1.2^\circ$ (black dotted) for the BAO feature, thereby delineating the angular extent of both features.\footnote{As discussed in detail in~\cite{Bashinsky:2000uh}, the BAO feature does not manifest itself as a peak in the angular correlation function $C(\theta)$. Instead, its wave front can be identified as an edge feature similar to the one in the upper right panel. }   

In multipole space, these two acoustic/sound horizon scales correspond approximately to
$\ell_{A,\mathrm{dec}} \sim 300$ and 
$\ell_{\mathrm{eq}} \sim 600$, which roughly correspond (up to a phase shift) to the first standard acoustic peak and to the pronounced dip within the blue-shaded band associated with the DAO position-space feature. The broadband effect, on the other hand, is visible as the overall suppression of power compared to the SIDR reference cosmology (with $f_\mathrm{idm}=0$). As discussed in more detail in \cite{Garny:2025kqj}, it arises because the radiation pressure exerted on dark matter leads to an excess decay of the gravitational potential (off-setting the effect of an increased $\omega_\mathrm{m}$). A feature at $0.6^o$ ($l\approx 600$) is well within the Planck regime, and not in the regime where ground based experiments like ACT and SPT are expected to improve the constraints. For more discussion of this aspect, see also section \ref{discussion}. 

These numerical values should be regarded as order-of-magnitude estimates. A precise characterization of the DAO requires the full Boltzmann analysis presented in the next section.

\section{Parameter Constraints and DAO Targets}\label{results}

\begin{figure}[t]
  \centering
    \includegraphics[width=0.48\textwidth]{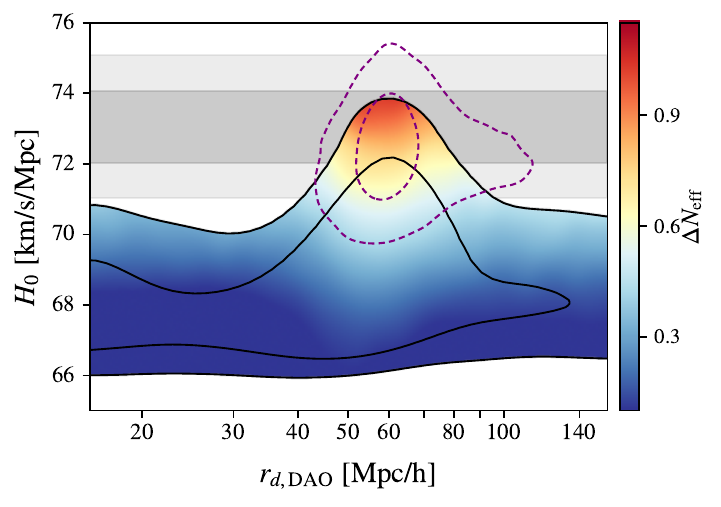}\hfill
  \includegraphics[width=0.48\textwidth]{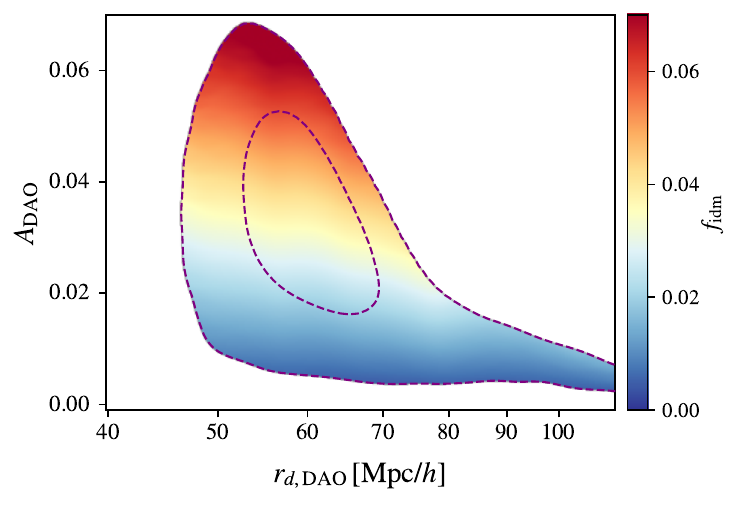}
    \vspace{-2mm}
\includegraphics[width=0.40\textwidth]{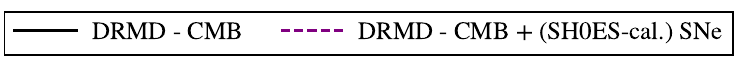}
\vspace{-4mm}
  \caption{\footnotesize Marginalized 2D posterior from our analysis using CMB data (black) and CMB combined with calibrated supernova (purple dashed). The inner and outer contours correspond to the 68\% and 95\% confidence levels. \textit{Left:} The SH$0$ES determination of $H_0$ is shown as the gray band~\cite{Riess:2021jrx}. Within DRMD, Planck 2018 becomes fully consistent with SH$0$ES. Marginalization over $r_{d,\mathrm{DAO}}$ induces strong projection effects due to parameter degeneracies, which are largely broken by including calibrated supernova data. \textit{Right:} Targets for future searches: DRMD predicts a percent-level DAO amplitude at dark drag horizon scales of approximately $60\, \mathrm{Mpc}/h$.}
  \label{fig:2D_posteriors}
\end{figure}

To obtain a quantitative prediction for the DAO feature, we perform a cosmological parameter inference using the publicly available \texttt{DRMD-CLASS} code~\cite{Garny:2025kqj}, which is based on the Cosmic Linear Anisotropy Solving System \texttt{CLASS}~\cite{Blas:2011rf}. We consider an extension of $\Lambda$CDM that introduces three additional parameters: the amount of self-interacting dark radiation $\Delta N_\mathrm{eff}$ (defined in~\eqref{eq:N_eff_2}), the fraction of interacting dark matter $f_\mathrm{idm}$ (defined in~\eqref{eq:f_idm_2}), and the redshift $z_\mathrm{stop}$ at which the exponential suppression of the drag rate sets in (defined in~\eqref{eq:z_stop}). 

The decoupling redshift $z_\mathrm{dec}$ is treated as a derived parameter using Eq.~\eqref{eq:z_dec}. Likewise, the DAO drag horizon $r_{d,\mathrm{DAO}}$ is derived from Eq.~\eqref{eq:r_D}. The initial interaction strength is fixed to an arbitrary value corresponding to initial tight coupling, $(\mathcal{G}/\mathcal{H})_\mathrm{ini} = 10^{7}$. As discussed in~\cite{Garny:2025kqj} (see also Eq.~\eqref{eq:G_ini}), for a given $z_\mathrm{dec}$ this parameter is degenerate with $z_\mathrm{stop}$, and including it in the sampling would just introduce a flat direction. In the current implementation we set $w_\mathrm{DR} = c_s^2 = 1/3$, neglecting mass-threshold effects, and assume a mechanism such as Hot NEDE to generate a sizeable $\Delta N_\mathrm{eff}$ after BBN~\cite{Garny:2024ums}. We accordingly use the Helium abundance prediction from BBN that applies for a $\Lambda$CDM cosmology with $\Delta N_\mathrm{eff}=0$.

For the parameter inference we use the general-purpose Bayesian analysis framework \texttt{Cobaya}~\cite{Torrado:2020dgo}. We impose flat priors
\begin{equation}
\Delta N_\mathrm{eff} \in [0,3], \qquad 
f_\mathrm{idm} \in [0,1], \qquad 
\log_{10}(z_\mathrm{stop}) \in [3,5.3],
\end{equation}
while adopting standard priors for the six $\Lambda$CDM parameters. We further assume one massive neutrino with $m_3 = 0.06\,\mathrm{eV}$ and temperature $T_3 = 0.716\,T_\gamma$.

We perform two Metropolis-Hastings Markov Chain Monte Carlo (MCMC) analyses. The baseline run (performed for  $\Lambda$CDM and DRMD) includes Planck 2018 CMB anisotropy measurements~\cite{Planck:2018vyg,Planck:2018lbu}, incorporating high-$\ell$ TT+TE+EE spectra, low-$\ell$ TT+EE spectra, and CMB lensing. The extended run additionally includes the Pantheon+ supernova sample~\cite{Brout:2022vxf}, calibrated using the SH$0$ES prior on the absolute magnitude $M$~\cite{Riess:2021jrx} (amounting to $H_0= 73.04 \pm 1.04\,\mathrm{km/s/Mpc}$ for late-time $\Lambda$CDM cosmologies). We quantify the residual Hubble tension using the posterior distribution of the difference $\Delta \equiv H_0^{\mathrm{CMB}} - H_0^{\mathrm{SH0ES}}$, obtained by convolving the marginalized one-dimensional posteriors of the two measurements (SH$0$ES and CMB only). The two-sided posterior tail probability for $\Delta=0$ is then converted into an equivalent Gaussian significance, which we quote as the residual tension in units of $\sigma$.

This setup follows a two-step logic. The baseline run demonstrates that DRMD reconciles Planck and SH$0$ES measurements. The extended run, which provides tighter parameter constraints, is then used to infer and predict the properties of the DAO feature.

The MCMC results are summarized in Tab.~\ref{tab:MCMC} and Fig.~\ref{fig:2D_posteriors}. 
The baseline DRMD run reduces the Hubble tension to the $2\sigma$ level, despite being affected by strong projection effects. 
This is evident in the left panel of Fig.~\ref{fig:2D_posteriors}, which shows a clear overlap of the 95\% credible regions between the SH$0$ES band (gray) and the DRMD posterior (color-coded).
Importantly, this overlap only occurs for dark drag-horizon scales in the range $40\,\mathrm{Mpc}/h < r_{d,\mathrm{DAO}} < 80\,\mathrm{Mpc}/h$.
Outside this interval, the two datasets become clearly incompatible. 
This demonstrates that the presence of the DAO feature is essential for making the CMB compatible with large values of $\Delta N_\mathrm{eff}$ and thus resolving the Hubble tension within the DRMD model.

At the same time, marginalizing over $r_{d,\mathrm{DAO}}$ (or equivalently over $\log_{10}(z_\mathrm{stop})$) induces a pronounced projection effect (commonly encountered in extensions of $\Lambda$CDM~\cite{Herold:2021ksg,Holm:2022kkd,Cruz:2023cxy}). 
This results in an apparently low marginalized value, $H_0 = 68.7^{+0.9}_{-1.8}\,\mathrm{km/s/Mpc}$. In principle, a profile-likelihood analysis can mitigate such projection effects. 
Indeed, for a closely related dataset combination, Ref.~\cite{Garny:2025kqj} reported a residual $1.4\sigma$ tension using the \textit{difference of the maximum a posteriori} (DMAP) estimator. 
Here, however, we restrict ourselves to the Bayesian analysis, which, using the method described above, results in a mild residual tension of $2.3 \sigma$ despite the presence of strong  projection effects. This is then further supported by the 2D posterior shown in Fig.~\ref{fig:2D_posteriors}, which similarly demonstrates that DRMD renders the two datasets statistically compatible. 

As expected, including the SH$0$ES-calibrated supernova data selects the region of parameter space in which the DAO horizon scale lies within the previously identified narrow interval and $\Delta N_\mathrm{eff}$ takes on sizeable values (purple dashed contour). 

For this dataset combination, we perform a more detailed characterization of the DAO signal beyond its horizon scale in a dedicated post-processing step. To be specific, we decompose the linear matter power spectrum as
\begin{equation}
P_\mathrm{lin}(k) = P_\mathrm{smooth}(k) + P_\mathrm{BAO}(k) + P_\mathrm{DAO}(k).
\end{equation}
An example of this decomposition is shown in Fig.~\ref{fig:DAO_BAO_split}, where the BAO and DAO contributions are displayed as the green dash-dotted and blue dashed curves, respectively.
The decomposition is obtained by performing, for each DRMD cosmology, a corresponding SIDR reference run with identical background evolution but no interactions between dark matter and dark radiation. In practice, this is implemented by setting $f_\mathrm{idm}=0$ while keeping the total dark matter abundance fixed. We then apply a smooth--wiggly split to both cosmologies using a Gaussian smoothing kernel, supplemented by a low-order polynomial correction to ensure an accurate broadband fit.\footnote{In practice, we apply a Gaussian smoothing in $\ln k$ with $\Delta\ln k = 0.25$, plus a cubic $\ln k$ polynomial correction fit to $\ln(P/P_{\rm smooth})$ over $k \in [0.001,5.5]\,h\,\mathrm{Mpc}^{-1}$.} For the DRMD cosmology, the oscillatory component contains both BAO and DAO contributions, whereas for the SIDR reference model it contains only BAO. This procedure therefore allows us to isolate and disentangle the DAO signal from the standard BAO feature.

The validity of this procedure is most directly assessed in position space. 
To this end, we consider the two-point correlation function,
\begin{equation}
\xi(r)
= \frac{1}{2\pi^2} \int_0^\infty dk \, k^2 P_{\mathrm{lin}}(k)\, \frac{\sin(kr)}{kr} \, ,
\label{eq:xi_of_r}
\end{equation}
which directly relates oscillatory features in $P_{\mathrm{lin}}(k)$ to localized structures in real space.
Applying this transform separately to $P_\mathrm{BAO}$ and $P_\mathrm{DAO}$ for our best-fit DRMD cosmology yields distinct peaks at 
$r \approx 100\,\mathrm{Mpc}/h$ and $r \approx 60\,\mathrm{Mpc}/h$, respectively, as shown in the lower-right panel of Fig.~\ref{fig:DAO_BAO_split}. 
Importantly, the peaks do not exhibit noticeable leakage: the BAO component does not generate a spurious feature at the DAO scale, and vice versa. 
This clean separation in configuration space confirms the robustness of our decomposition procedure.

This analysis allows us to quantify the amplitudes of the oscillatory features. 
We define
\begin{align}
A_\mathrm{DAO} &= \max\!\left(\frac{|P_\mathrm{DAO}|}{P_\mathrm{smooth}}\right) \, ,
\qquad
A_\mathrm{BAO} = \max\!\left(\frac{|P_\mathrm{BAO}|}{P_\mathrm{smooth}}\right) \, ,
\end{align}
where the maxima are taken over  $k \in [0.01,2]\,h\,\mathrm{Mpc}^{-1}$.

This procedure admits a quantitative prediction of the DAO feature in terms of its drag horizon and amplitude. As the main result of this work, we obtain
\begin{equation}\label{eq:prediction}
r_{d,\mathrm{DAO}} \in [53.7,\,64.8]\,\mathrm{Mpc}/h\,,
\qquad
A_\mathrm{DAO} = 0.031^{+0.014}_{-0.011} 
\quad (68\%\,\mathrm{C.I.}) \, .
\end{equation}
These predictions for the properties of a DAO feature in the matter power spectrum obtained from the resolution of the Hubble tension within the DRMD model based on only Planck CMB and (SH$0$ES-calibrated) supernova data are illustrated by the 2D posterior contours in the right panel of Fig.~\ref{fig:2D_posteriors} and can be used for a targeted feature search in large-scale structure data. As a simple consistency check, the color-coding in Fig.~\ref{fig:2D_posteriors} indicates that larger values of $A_\mathrm{DAO}$ correspond to larger values of $f_\mathrm{idm}$, as expected. 
For comparison, the considerably tighter BAO constraints read
\begin{equation}
r_{d,\mathrm{BAO}} = 100.1 \pm 1.1\,\mathrm{Mpc}/h,
\qquad
A_\mathrm{BAO} =0.0531^{+0.0013}_{-0.0015}
\quad (68\%\,\mathrm{C.I.}) \, .
\end{equation}
In other words, the DAO amplitude can be substantially smaller than its BAO counterpart. However, if the model is to play a significant role in addressing the Hubble tension, $A_\mathrm{DAO}$ should not fall far below the percent level.

In Fig.~\ref{fig:r_DAO_1D}, we display the one-dimensional posterior for the drag horizon $r_{d,\mathrm{DAO}}$ (purple dashed) and compare it with results from earlier analyses. 
We find excellent agreement with the constraints reported in Ref.~\cite{Garny:2025kqj} (blue dotted), which considered a four-parameter version of the DRMD model (featuring an additional degeneracy direction associated with $(\mathcal{G}/\mathcal{H})_\mathrm{ini}$) and included DESI DR2 BAO data.\footnote{The one-dimensional posterior shown here was obtained by post-processing the MCMC chains of Ref.~\cite{Garny:2025kqj} to include $r_{d,\mathrm{DAO}}$ as a derived parameter using Eq.~\eqref{eq:r_D}.} 
Remarkably, our result is also compatible with the ``negative branch'' identified in the DAO bias analysis of Ref.~\cite{Garny:2025szk} in the context of the DESI anomaly. We stress that the present determination constitutes an entirely independent inference of the DAO signal. Thus the match of the CMB/SH0ES preference for a DAO scale of order $60$ Mpc$/h$ related to resolving the Hubble tension and the independent DESI DR2 preference related to the DESI anomaly constitutes a non-trivial concordance within DRMD.

Overall, if the Hubble tension is resolved through new decoupling physics as captured by the DRMD model, the above values constitute a falsifiable signature of this scenario that can be looked for in current and future large-scale structure surveys.

\section{Discussion and Outlook}\label{discussion}

Both the Hubble tension and the DESI anomaly challenge the $\Lambda$CDM model as a complete description of our Universe. Among them, the Hubble tension, a discrepancy between the value of $H_0$ inferred from CMB fits within $\Lambda$CDM and the locally measured value from supernova observations, currently provides the most statistically significant evidence for physics beyond $\Lambda$CDM.

The DRMD model is a three-parameter extension of $\Lambda$CDM, featuring a component of interacting dark matter, which decouples from a component of self-interacting dark radiation around matter-radiation equality. We have shown that DRMD can resolve the Hubble tension when fitted only to Planck 2018 CMB data. In this case, the Hubble tension is reduced to $2.3\sigma$, compared to more than $5\sigma$ in $\Lambda$CDM. Crucially, a decoupling of interacting dark matter and self-interacting dark radiation around matter-radiation equality predicts the existence of DAO on an acoustic scale of roughly half the BAO scale. The regime of DRMD fitted to Planck 2018 data, which reduces the Hubble tension below $3\sigma$ shows a clear preference for a DAO with these features.

Once we complement the data analysis with SH0ES-calibrated supernova data, this preference sharpens into the prediction quoted in \eqref{eq:prediction}, independently of large-scale structure surveys such as DESI. We note that the posterior for $r_{d,\mathrm{DAO}}$ is markedly non-Gaussian and extends to large values of $r_{d,\mathrm{DAO}}$, including values exceeding $r_{d,\mathrm{BAO}}$, always at non-zero $A_\mathrm{DAO}$.

Independently of DRMD and the Hubble tension, in previous work \cite{Garny:2025szk} we investigated whether the anomaly in the DESI DR2 data, typically interpreted in terms of evolving dark energy, can instead be explained by the presence of a DAO peak which biases the determination of the BAO peak position. We showed that DAO can indeed explain the anomaly and are preferred over the null hypothesis of no DAO with more than five units of $\chi^2$, with a preferred position and amplitude closely overlapping with the DAO properties found in this work (independently of DESI and other large-scale structure data using only CMB and supernova data). The presence of DAO provides an arguably more convincing interpretation of the DESI DR2 data than evolving dark energy, the interpretation preferred by the DESI collaboration itself, as evolving dark energy cannot obviously be related to the Hubble tension. From an Occam's razor standpoint, we therefore believe the DAO interpretation of the DESI anomaly, which connects both problems, is a more credible explanation.

The predictions of the position and amplitude of the DAO obtained from fitting DRMD to the CMB and (SH$0$ES-calibrated) supernova data, independent of BAO and large-scale structure data, provides a target for present and future large-scale structure surveys like DESI \cite{DESI:2023dwi}, Euclid \cite{Euclid:2024yrr}, and the Roman Space Telescope \cite{Spergel:2015sza}. The evidence for the DAO from the bias analysis of the DESI DR2 data provides a first preliminary and independent validation from large-scale structure surveys. Future full-shape analyses of the DESI data as well as large-scale structure data will be able to constrain more decisively whether DAO arising within the DRMD model provide an explanation of the Hubble tension and the DESI anomaly.

The DAO feature predicted by the DRMD model in the matter power spectrum leaves only a much weaker imprint in the CMB angular power spectrum, since the DM-DR interaction directly impacts matter perturbations but only indirectly affects CMB photons. Nevertheless, as is well-known, the extra relativistic dark fluid component also leads to a broadband effect on the CMB on small angular scales. Therefore,
another important constraint on DRMD comes from new CMB observations extending to smaller scales and yielding higher quality polarization data than Planck, such as ACT \cite{AtacamaCosmologyTelescope:2025nti} or SPT \cite{SPT-3G:2025bzu}. The latest results from ACT and SPT are painting a complicated picture. While the ACT constraints on (free-streaming) $\Delta N_\textrm{eff}$ are tighter than the constraints from Planck, the SPT constraints on $\Delta N_\textrm{eff}$ are more loose. The stronger ACT constraints arise because ACT observes more power on small scales in the CMB, which is contrary to the extra radiation-damping one would observe with a sizeable $\Delta N_\textrm{eff}\gtrsim 0.5$ as predicted in DRMD. One could object that DRMD relies on self-interacting (rather than free-streaming) radiation and features additional interactions with dark matter not present in simple $\Lambda$CDM + $N_\mathrm{eff}$ cosmologies, but first results in the literature indicate that this is still not sufficient to reach large enough values of $H_0$~\cite{Cvetko:2025kda} to resolve the Hubble tension. Therefore, if the DRMD model is the correct solution to the Hubble tension, there are two possible explanations of the ACT data. One possibility is a residual foreground effect in ACT data, as the scales where ACT observes extra power is exactly the scales where foreground effects quickly become sizeable, diminishing the signal-to-noise ratio. While the ACT collaboration has made significant and impressive efforts to ensure that residual foreground effects are not driving their stronger constraints, this possibility needs yet to be tested further. Another possibility is to mildly extend the cosmological model to make it consistent with ACT data, while still resolving the Hubble tension \cite{workinprogress}. We have shown that the DAO feature is a localized feature in real angular space located at approximately $0.3^o$, which corresponds to around $l\approx 600$ in multipole space, which is well within the Planck regime. We therefore do not expect that any mild parameter extensions addressing the issue with ACT data on small scales will affect the DAO prediction. The predictions presented here can thus be considered as a robust prediction of any model, within the DRMD class of models, solving the Hubble tension and agreeing with CMB observations on angular scales dominated by Planck, while potentially allowing for additional physics affecting only the broadband shape on very small angular scales.

\newpage

\subsection*{Acknowledgements}
We acknowledge Shi-Fan (Stephen) Chen, Ricardo Z.~Ferreira, Laura Herold, Mikhail Ivanov, and Benjamin Wallisch for useful conversations. MG acknowledges support by the Excellence Cluster ORIGINS, which is funded by the Deutsche Forschungsgemeinschaft (DFG, German Research Foundation) under Germany’s Excellence Strategy – EXC-2094 - 390783311 as well as the DFG Collaborative Research Centre ``Neutrinos and Dark
Matter in Astro- and Particle Physics'' (SFB 1258). FN was supported by VR Starting Grant 2022-03160 of the Swedish Research Council. MSS acknowledges Nordita for their kind hospitality through the Nordita corresponding fellow program.

\begin{table}[h!]
\centering
\fontsize{7.5}{8.5}\selectfont
\setlength{\tabcolsep}{8pt} 

{
\begin{tabular}{c|c|c|c}
 &\textbf{$\Lambda$CDM}&\textbf{DRMD}&\makecell{\textbf{DRMD}\\ + (SH$0$ES-cal.)~SN}\\
\hline
\rule{0pt}{4ex}\textbf{Parameter}& \makecell{$68 \%$ limits  \\(best-fit)}  & \makecell{$68 \%$ limits  \\(best-fit)}  & \makecell{$68 \%$ limits  \\(best-fit)} \\
\hline
\hline
\rule{0pt}{3ex}\makecell{$\omega_b$} & \makecell{$0.02236\pm 0.00015$\\$(0.02240)$} & \makecell{$0.02250^{+0.00018}_{-0.00027}$\\$(0.02255)$} & \makecell{$0.02311\pm 0.00018$\\$(0.02318)$}
\\
\hline
\rule{0pt}{3ex}\makecell{$\omega_{\mathrm{cdm}}$} & \makecell{$0.1199\pm 0.0012$\\$(0.1198)$} & \makecell{$0.1241^{+0.0019}_{-0.0053}$\\$(0.1258)$} & \makecell{$0.1375\pm 0.0046$\\$(0.1382)$}
\\
\hline
\rule{0pt}{3ex}\makecell{$H_0\,\mathrm{[km/s/Mpc]}$} & \makecell{$67.36\pm 0.54$\\$(67.43)$} & \makecell{$68.32^{+0.71}_{-1.7}$\\$(68.23)$} & \makecell{$72.40\pm 0.91$\\$(72.50)$}
\\
\hline
\rule{0pt}{3ex}\makecell{$\ln(10^{10} A_s)$} & \makecell{$3.044\pm 0.015$\\$(3.047)$} & \makecell{$3.045\pm 0.015$\\$(3.042)$} & \makecell{$3.047\pm 0.015$\\$(3.051)$}
\\
\hline
\rule{0pt}{3ex}\makecell{$n_\mathrm{s}$} & \makecell{$0.9646\pm 0.0041$\\$(0.9659)$} & \makecell{$0.9674^{+0.0047}_{-0.0062}$\\$(0.9688)$} & \makecell{$0.9778\pm 0.0048$\\$(0.9798)$}
\\
\hline
\rule{0pt}{3ex}\makecell{$\tau_{\mathrm{reio}}$} & \makecell{$0.0541\pm 0.0076$\\$(0.0564)$} & \makecell{$0.0543\pm 0.0075$\\$(0.0507)$} & \makecell{$0.0574\pm 0.0075$\\$(0.0581)$}
\\
\hline
\rule{0pt}{3ex}\makecell{$\Delta N_\mathrm{eff}$} & -- & \makecell{$< 0.590$\\$(0.217)$} & \makecell{$0.85\pm 0.17$\\$(0.87)$}
\\
\hline
\rule{0pt}{3ex}\makecell{$\log_{10}(z_\mathrm{dec})$} & -- & \makecell{$3.32^{+0.77}_{-0.16}$\\$(3.34)$} & \makecell{$3.316^{+0.075}_{-0.024}$\\$(3.350)$}
\\
\hline
\rule{0pt}{3ex}\makecell{$f_\mathrm{idm}$} & -- & \makecell{$< 0.702$\\$(0.020)$} & \makecell{$0.035^{+0.015}_{-0.013}$\\$(0.039)$}
\\
\hline
\hline
\rule{0pt}{3ex}\makecell{$\Omega_m$} & \makecell{$0.3151\pm 0.0074$\\$(0.3127)$} & \makecell{$0.3156\pm 0.0093$\\$(0.3185)$} & \makecell{$0.3076\pm 0.0082$\\$(0.3070)$}
\\
\hline
\rule{0pt}{3ex}\makecell{$r_{d,\mathrm{BAO}}\, \mathrm{[Mpc/h]}$} & \makecell{$99.11\pm 0.93$} & \makecell{$99.1\pm 1.2$\\$(98.5)$} & \makecell{$100.1\pm 1.1$\\$(100.0)$}
\\
\hline
\rule{0pt}{3ex}\makecell{$r_{d,\mathrm{DAO}}\, \mathrm{[Mpc/h]}$} & -- & \makecell{$57^{+9}_{-60}$\\$(55.154)$} & \makecell{$\left[53.7,\,64.8\right]$\\$(58.6)$}
\\
\hline
\rule{0pt}{3ex}\makecell{$(r_{d, \mathrm{DAO}}-r_{d,\mathrm{BAO}})/r_{d,\mathrm{BAO}}$} & -- & -- & \makecell{$\left[-0.460,\,-0.352\right]$}
\\
\hline
\rule{0pt}{3ex}\makecell{$A_\mathrm{DAO}$} & -- & -- & \makecell{$0.031^{+0.014}_{-0.011}$\\$(0.036)$}
\\
\hline
\rule{0pt}{3ex}\makecell{$A_\mathrm{BAO}$} & -- & -- & \makecell{$0.0531^{+0.0013}_{-0.0015}$\\$(0.0538)$}
\\
\hline
\rule{0pt}{3ex}\makecell{$\tilde A = A_\mathrm{DAO}/A_\mathrm{BAO}$} & -- & -- & \makecell{$0.59^{+0.25}_{-0.20}$\\$(0.67)$}\\
\hline
{$H_0$ tension (Bayesian) }&$5\sigma$  & $2.3\sigma$ & --\\
\hline
{$\Delta \chi^2$ (rel.~to $\Lambda$CDM)  }&$0$  & -1.9 & -31.7\\
\end{tabular}
} 

\caption{\label{tab:MCMC}\footnotesize Posterior means and 68\% credible intervals for parameters inferred from the baseline analysis under $\Lambda$CDM and DRMD (left and middle columns), and from the extended analysis including calibrated supernova (SN) data (right column). One-sided constraints are quoted at 95\% confidence. In cases where the posterior mean lies outside the 68\% credible interval due to extended one-sided tails, we report only the credible interval (centred on the median).
 }
\end{table}

\newpage

\bibliography{Ref}
\end{document}